\documentclass[12pt,preprint]{aastex}
\usepackage{epsfig,rotating,amsmath,subfigure}
\usepackage{verbatim,longtable}
\usepackage{natbib,lscape}

\shorttitle{IR detections of small HI clouds}
\shortauthors{Saul et al.}
\slugcomment{Accepted to MNRAS}

\def\HI      {\ion{H}{1}~}
\def\vlsr    {V$_{\rm LSR}$}
\def\kms   {~km~s$^{-1}$}
\def\eg         {{\it e.g.,\ }}
\def\dgrunits {$\times~10^{-20}$ MJy Sr$^{-1}$ cm$^{-2}$}

\begin{document}

\title{Dust-to-Gas Ratios of the GALFA-\HI Compact Cloud Catalog}

\author{Destry R. Saul\altaffilmark{1}, J. E. G. Peek\altaffilmark{1,2}, M. E. Putman\altaffilmark{1}}
\altaffiltext{1}{Department of Astronomy, Columbia University, New York, NY 10027}
\altaffiltext{2}{Hubble Fellow}

\begin{abstract}
We have searched for infrared dust emission from subsets of compact, Galactic neutral hydrogen clouds, with
the purpose of looking for dust in high-velocity clouds, identifying low-velocity halo clouds, and investigating
the cloud populations defined in the GALFA-HI Compact Cloud Catalog. We do not detect dust emission from
high-velocity clouds. The lack of dust emission from a group of low-velocity clouds supports the claim
that they are low-velocity halo clouds. We detect dust in the remaining low-velocity clouds, indicating a Galactic origin, with a significantly greater 
dust-to-gas ratio for clouds with linewidths near 15~\kms.
We propose that this is due to dust associated with ionized gas.
\end{abstract}

\keywords{ISM: clouds, ISM: structure, ISM: dust, Galaxy: halo, infrared:ISM, radio lines:ISM}


\section{Introduction}

The Galactic dust-to-gas emission ratio (DGR) has been observed to be uniform for moderate \HI column densities ($10^{19}-10^{21}$ cm$^{-2}$) 
at low-velocities ($|$\vlsr$| < 90$\kms) and 100pc scales \citep{boulanger88,jones95}, while the DGR of high-velocity gas is significantly reduced, 
if not zero \citep{wakker86,miville05,peek09,planck11}. In this paper, we measure the DGR using the ratio of Far-IR flux to \HI column density as traced
by 21cm luminosity, assuming optically thin gas.
The uniformity of the Galactic DGR is evidence for a well-mixed ISM \citep{boulanger85}. At high \HI column densities, the gas becomes dense enough to self-absorb and to
form molecular hydrogen . This reduces the observed 21cm emission which increases the measured DGR for dense regions \citep{blitz90,reach94,reach98,meyerdierks96,boulanger98,douglas07}. Because the Far-IR is tracing the
total hydrogen column and not just the \HI, the DGR can be used to detect the presence of 'dark gas' which we cannot detect in 21cm 
emission \citep{planck11_19}.

While high column density gas has an elevated DGR, high-velocity clouds (HVCs) with $|$\vlsr$| > 90$\kms ~have a reduced, if not zero, DGR
\citep{wakker86,miville05,peek09}. There are two main explanations for the low DGR, either HVCs have significantly less dust than Galactic gas, or
the dust they do have is not radiating sufficiently to be detectable. HVCs are known to have low metallicities, which correlates with
a low dust content \citep{wakker97}. HVCs are also known to be far from the Galactic disc, so the interstellar radiation field (ISRF) will
not heat the dust as it would the ISM.
\cite{peek09} used the reduced DGR of high-velocity clouds (HVCs) to search for their low-velocity analogs - low-velocity halo clouds (LHVCs).
LVHCs are an important component of the baryon cycle. 
If gas is cooling (or staying neutral) in the halo before accreting onto the Galactic disc, simulations show that
a significant quantity of gas could be at low radial velocities \citep{sommerLarsen06, peek08}. Other simulations have
shown that infalling HVCs could recool after disrupting in the halo \citep{heitsch09}. The recooled clouds would be 
observable at low radial velocities. Instead of accretion, neutral clouds could condense out of
the hot halo \citep{maller04}, but some simulations predict that neutral hydrogen clouds cannot form from linear perturbations
\citep{joung12}. Lastly, due to geometric effects, HVCs can have low radial velocities while still having high infall velocities.
Detected LVHCs could provide important constraints for models of halo processes (see \cite{putman12} for a 
review of gas in the Galactic halo). \cite{peek09} identified three candidate LVHCs; two with DGRs agreeing with zero, and one with a very low DGR.

Besides searching for LVHCs, the DGR of neutral hydrogen clouds can be used to probe the dust properties of gas that 
may be part of the Galactic fountain \citep{shapiro76}. Low and intermediate velocity clouds may be in the process of being 
ejected from the disc or accreting onto it.
Recent models by \cite{marasco12} show that supernova-driven accretion could be a major source of 
fuel for star formation. In this model, ejecta from supernovae provide the nonlinear perturbations necessary for cloud condensation.
Observations of ionized metals in the lower halo \citep{shull09,collins09,lehner12} agree with 
the supernova-driven model \citep{fraternali13} indicating that corresponding neutral clouds could be dust rich.

The GALFA-\HI Compact Cloud Catalog \citep{saul12}, hereafter GC3, is an ideal dataset to probe lower column densities, smaller size scales, and search
for dust in HVCs or a lack thereof in LVCs. 
By definition, the clouds in the GC3 are distinct from extended Galactic \HI emission and smaller than 20\arcmin ~in angular size.
The GC3 is separated into populations of clouds based on position, velocity, and linewidth distributions that Far-IR detections could support.
In this work we make two sets of measurements: the 100\micron/21cm and the 60\micron/21cm ratio in compact clouds. 
From previous observations, we expect that the 60\micron/21cm ratio will be a factor of four lower than the 100\micron/21cm ratio so we will use the 100\micron/21cm ratio to look for the presence of dust. 
Where we observe 60\micron ~emission we can explore the size or temperature of the dust grains as smaller grains will radiate more efficiently at shorter wavelengths.
Previous observations of the Galactic ISM have measured 100\micron/21cm DGRs of 0.5-2 \dgrunits~and 60\micron/21cm DGRs 
of 0.1-0.25 \dgrunits~\citep{boulanger88,jones95,reach98,planck11}.

We have searched for IR emission from a catalog of small neutral hydrogen clouds to look for answers to several questions: 
Can we detect dust in HVCs? 
Is the Galactic DGR constant at smaller scales and lower column densities than have been measured before?
Can we detect a population of LVHCs? 
Does the IR emission agree with the population definitions of \cite{saul12}?
This paper 
is organized as follows: we describe the data we use in \S \ref{sec:data}, in \S \ref{sec:methods} we detail the preprocessing of the data, the DGR detection method, and uncertainty
calculations. We report our results in \S \ref{sec:results}, and provide interpretation in \S \ref{sec:discussion}.

\section{Data}
\label{sec:data}

The clouds we are inspecting were identified in the GALFA-HI Compact Cloud Catalog \citep{saul12}.
The Galactic Arecibo L-Band Feed Array HI (GALFA-HI) survey is a large-scale neutral hydrogen survey that is being completed with the Arecibo telescope. 
The first data release was made publicly available in 2011 \citep{peek11}.
The beamsize of the GALFA-\HI survey is $\sim4\arcmin$. The clouds in the GC3 were identified by a machine-vision algorithm and visually confirmed.
They are limited to angular sizes of 4\arcmin ~to 20\arcmin ~and have typical FWHM linewidths of 3 - 30 \kms.
The catalog is separated into five populations based on position, velocity, and linewidth: High Velocity Clouds
(HVCs) have $|$\vlsr$|> 90$\kms ~and are associated with a previously observed HVC complexes as defined by \cite{peek07}, Galaxy Candidates (GCs) are the remaining clouds with $|$\vlsr$|> 90$\kms ~that are not associated with an HVC complex, Cold Low-Velocity Clouds (CLVCs) are clouds with $|$\vlsr$|< 90$\kms ~and 
linewidths less than 15 \kms, Warm Low-Velocity Clouds (WLVCs) are clouds with $|$\vlsr$|< 90$\kms ~and
linewidths greater than 15 \kms ~{\it excluding} the Warm, Positive-Velocity Clouds in the third Galactic quadrant (+WQ3) which are clouds with $90 >$\vlsr $> 0 $\kms, linewidths greater than 15 \kms, and 
$270 > l > 180$.

For the infrared observations we chose to use the IRIS \citep{miville05iris} reprocessing of IRAS \citep{beichman87}
because of the completeness of the sky coverage at 60 and 100\micron
~and that the IRIS resolution is 4.0\arcmin$\pm0.2$ and 4.3\arcmin$\pm0.2$ at 60 and 100\micron, respectively. 
Because the IR and \HI data have similar resolutions, we do not need to resample either dataset.
The IRIS noise level is $0.03\pm0.01$ and  $0.06\pm0.02$ MJy Sr$^{-1}$ at 60 and 100\micron, respectively. 

\section{Methods}
\label{sec:methods}

Relative to previous measurements of the DGR of Galactic clouds \citep[e.g.][]{miville05,peek09,planck11}, we are investigating simple objects, so 
there is no need for complicated detection algorithms.
We chose not to use the displacement map method presented by \cite{peek09}, or the  multiple-component fits used by \cite{miville05} and the Planck Collaboration (2011).
Instead, we preprocess the images to remove artifacts, point sources, and as much non-cloud emission as possible.
We then stack a subset of clouds using median statistics and fit a linear relation using a least-squares fit. We calculate the uncertainty in our measurements
using off-positions and a bootstrap analysis \citep{efron82}. This is discussed further in \S\ref{sec:error}.

\subsection{Preprocessing}
\label{sec:preprocessing}
\begin{figure}
\includegraphics[scale=0.8,angle=0]{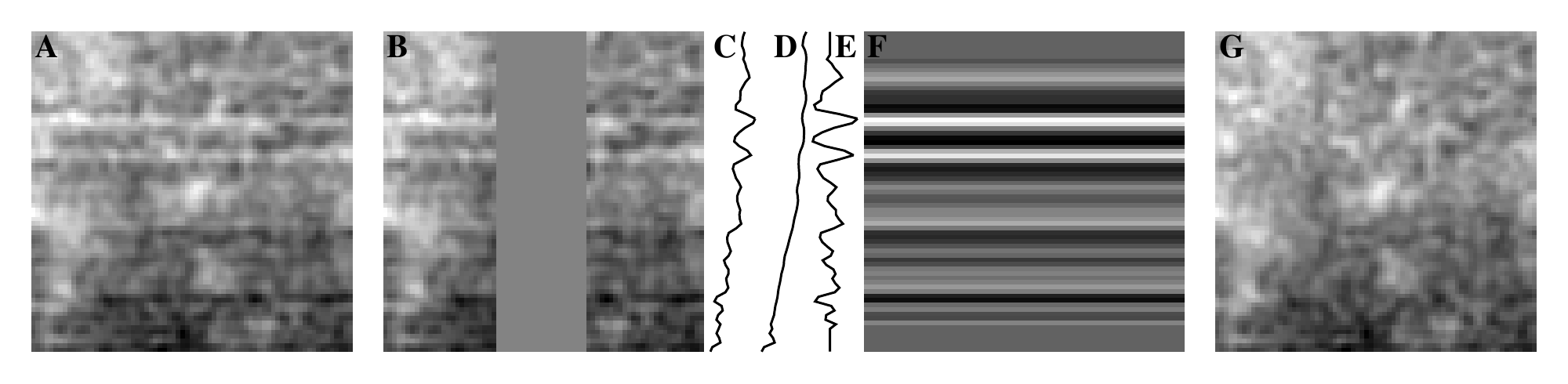}
\caption[A graphical example of our destriping procedure]{A graphical example of our destriping procedure.
{\bf A.} This is a raw, velocity-integrated \HI image centered on a compact cloud. The horizontal striping is due to residual baseline 
ripple variation in the GALFA-\HI survey.
{\bf B.} We blank the center columns of the image to remove the signal from the cloud.
{\bf C.} The average column is produced by summing parallel to the direction of striation.
{\bf D.} The striping is produced by variations in the baseline of the different passes of the ALFA seven-element beam 
and therefore must be smaller than 12\arcmin.  We smooth the average column (C) with a 12\arcmin ~boxcar window to isolate 
non-striation features.
{\bf E.} We subtract the non-striation features (D) from the average column (C) to isolate the the striation profile.
{\bf F.} We extrude the striation profile (E) into a two dimensional model.
{\bf G.} Subtracting the striation model (F) from the raw image (A) produces a nearly stripe-free image.}
\label{fig:destripe}
\end{figure}

The individual \HI images were created by integrating over a velocity range centered on the peak velocity of each cloud and covering twice the FWHM. 
The images were converted to units of cm$^{-2}$ assuming optically thin gas.
In regions that were observed in `drift' mode, there was visible striping in the images due to residual baseline ripple variation 
(for an explanation of this artifact, see \citealt{peek11}). To remove the stripes, we applied a 
2D filter to all of the data. The destriping algorithm is detailed in Figure \ref{fig:destripe}. 

For the Far-IR, we extracted images of each cloud from the IRIS data. We identified point sources with IDL's
implementation of DAOPHOT's FIND \citep{stetson87}. We removed the identified point sources
by fitting a 2D gaussian at the location returned by FIND, and subtracting the fit. There was no overdensity
of identified point sources at the locations of the compact clouds.

One difficulty in matching IR to \HI emission is that the IR images
contain emission from all dust along the line of sight, while the \HI images of the clouds show only a velocity-selected sample. Previous studies
have approached this problem in different ways. Early work, such as \cite{boulanger88} and \cite{wakker86}, did not attempt 
to separate the \HI data by velocity. \cite{peek09} used a sophisticated algorithm to pattern-match the \HI and IR images. 
\cite{miville05} and the Plank Collaboration (2011) both use multiple component fits to measure the DGRs of different velocity components.

We found that using the integrated low-velocity \HI data to
subtract the non-cloud IR emission introduced uncertainties of the same magnitude as the clouds themselves, so we chose not to use that technique. 
Instead, we take the median of a 4\arcmin ~wide annulus surrounding a 10\arcmin ~radius circle centered on the cloud and subtract that value from the IR image. This will remove the non-cloud
IR emission assuming the non-cloud emission varies slowly relative to the cloud size. We test this assumption by using off positions to estimate the uncertainty
in our detections. (see  \S \ref{sec:error} for a detailed discussion on errors and significance). We apply the same technique to the \HI images to 
subtract any large-scale emission or baseline offset.
\subsection{Stacking And Measurements}
\label{sec:measurements}
\begin{figure}
\includegraphics[scale=0.9,angle=0]{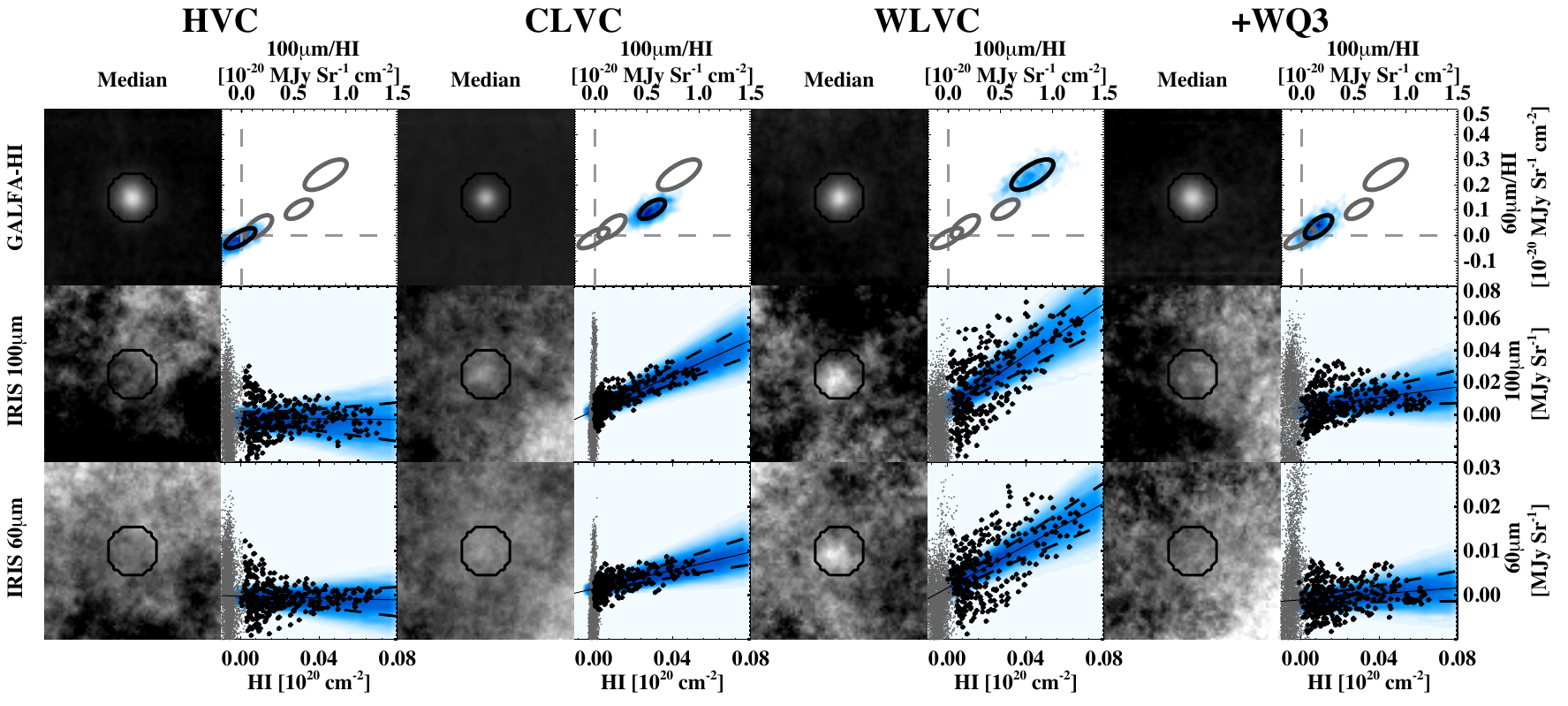}
\caption[IR and \HI median images and DGR measurements]{IR and \HI median images and DGR measurements for the four cloud populations: high velocity clouds (HVC), cold low-velocity clouds (CLVC)
warm low-velocity clouds (WLVC), and warm low positive velocity clouds in the third galactic quadrant (+WQ3). Cold clouds have FWHM line widths less
than 15 \kms and warm clouds have FWHM line widths greater than 15 \kms. The median stacked image for each population is shown for 21 cm \HI (upper left), 100\micron (center left), and 60\micron (lower left) , with the same scale for all populations. 
The area used for the DGR fit is outlined by the black circle in each image. The fit results for the DGR are shown for 100\micron (center right) and 60\micron 
(lower right). The pixel values used in the fit are black while the pixel values for the rest of the images are light grey. The measured DGRs are plotted with a solid
black line over the results of the 1000 iteration bootstrap error analysis, shown in blue. 
The $1\sigma$ uncertainties are plotted with dashed lines. The upper right panel
plots the results of the bootstrap calculation in blue, over plotted with the 1-sigma covariance ellipse of each population. The DGR of the population of the plots below is shown in black. Note the non-detections of the HVC and +WQ3 
populations and detections of the CLVC and WLVC populations.}
\label{fig:main}
\end{figure}

We stack the clouds in a given set by taking the median value of each pixel location. 
We chose to use the median instead of the mean because it is less affected by outliers \citep{gott01,white07}. 
To measure the median DGR of a set of clouds, we begin by describing the Far-IR image for one image:

\begin{equation}
I_\lambda = D_{\lambda,C} N_C + \sum_{i}{D_{\lambda,i}N_i} + K_\lambda
\end{equation}

\noindent where $I_\lambda$ is the IR intensity at wavelength $\lambda$, $D_{\lambda,C}$ is the DGR of cloud $C$ at 
wavelength $\lambda$, $N_C$ is the \HI column density of cloud $C$, $K_\lambda$ is a constant term, and the sum term represents all of the remaining non-cloud
gas along the line of sight. The non-cloud term is the greatest source of uncertainty
as it is difficult to disentangle the IR emission from multiple clouds. We found that attempting 
to remove the non-cloud emission from the IR images introduced addition uncertainty on a scale similar to the clouds. We utilize the 
small size of the clouds relative to the non-cloud emission, the annulus background subtraction, and the median
stacking to set the non-cloud sum term to zero, leaving a simpler linear equation

\begin{equation}
\tilde{I}_\lambda = \tilde{D}_{\lambda,C} \tilde{N}_C + \tilde{K}_\lambda,
\end{equation}

\noindent where $\tilde{I}$ denotes the median of $I$. We solve this simpler equation using standard least-squares with the MPFIT package \citep{mpfit}. 
We include pixels within 10\arcmin ~of cloud center in the fit.
Figure \ref{fig:main} displays the values both inside and outside the fitting area with the result of the
fits for both IR bands for the populations of the GC3.

\subsection{Errors and Significance}
\label{sec:error}

There are two major sources of uncertainty in our measurements - non-cloud emission and variation in the DGR. 
Firstly, the IR emission from dust that is not associated with the compact clouds has structure 
and could be falsely identified as emission from the clouds. To investigate the effect of non-cloud emission, 
we take a grid of 100 off-source positions, separated by one degree in both right ascension and declination, centered on each cloud. 
For each stacked image, we calculate the DGR for each 
of the 100 off positions and use the variation as the uncertainty introduced by the non-cloud IR emission. This technique also tests
for bias introduced by the annulus subtraction. The mean values of the off position measurements were consistently near zero, indicating no significant bias. The uncertainty of our
measurements is dominated by the non-cloud emission. Empirically, we found that we were unable to reduce the uncertainty below
$\sim0.1$\dgrunits at 100\micron ~and $\sim0.04$\dgrunits at 60\micron.

The other uncertainty in our measurements is that each cloud could have a different DGR, thus each measurement is not a 
sample of the same DGR, but a sample from the DGR distribution. 
To calculate errors in the stacked image we use the bootstrap method \citep{efron82}. For each set of clouds, we resample with 
replacement and calculate the DGR many times (n= 1000) until there are a sufficient number of measurements to calculate uncertainties from.

The bootstrap error calculation includes the variance in both the DGRs of the clouds and the non-cloud emission, as each IR image
is at a different position (it is physically possible to have two \HI clouds at the same location on the sky and at different velocities, but 
we do not have such a case). Thus, as we expect, the bootstrap errors are larger than the off position uncertainty for detections
and converging for non detections. 
We report the bootstrap results as our final uncertainties.

\section{Results}
\label{sec:results}
We begin our analysis by first investigating the HVC population, both as a whole, and separated by
complex. We then measure the DGR for the +WQ3 population that was proposed to be a low-velocity
halo cloud complex by \cite{saul12}. Finally, we search for significant trends in the DGR of the low
velocity clouds by separating them into sub-populations. 

\subsection{High-Velocity Clouds}
\label{sec:results_hvcs}


We do not detect any dust emission at the $3\sigma$ level for the HVC population as a whole or in any particular complex.
The total HVC stack is show in the upper left pane of Figure \ref{fig:main}.
When the clouds were stacked by complex, there were two $2\sigma$ detections: Complex L at 100\micron ~and Complex WB at 60\micron.

The complex L measurement is 2.34 $\pm$ 1.00 \dgrunits (100\micron) and 0.56 $\pm$ 0.30 \dgrunits (60\micron) from a subset of eight clouds.
We approach these measurements with caution due to the small number of clouds because
other subsets with the same sample size have larger error bars. It appears that these clouds are in regions of lower non-cloud IR variation.
That said, Complex L is particularly bright in H$\alpha$ ~emission for an HVC complex and has elevated levels of [\ion{N}{2}]/H$\alpha$ \citep{haffner05}.
Measurements near Complex have shown super-solar metallicities \citep{zech08}. This suggests that Complex L is more similar to Galactic fountain clouds than other HVCs.

The complex WB measurement is 0.31 $\pm$ 0.23 \dgrunits (100\micron) and  0.21 $\pm$ 0.09 \dgrunits (60\micron) from a subset of 53 clouds. The 100\micron ~measurement 
agrees with zero, while the 60\micron ~value differs by $2\sigma$. With 53 clouds, the uncertainties are more reliable than the complex L results,
but a 60/100\micron ~ratio of $\sim0.7$ is extreme and makes this $2\sigma$ detection questionable. This complex is located in the halo (D $\sim$ 10kpc) and is
part of a larger group of scattered clouds in this region of the sky \citep{thom06}.

\subsection{Warm, Positive-Velocity, Third Quadrant Clouds} 
\label{sec:results:q3clouds}
The measured DGR of the +WQ3 clouds is 0.16 $\pm$ 0.13 \dgrunits (100\micron) and 0.03 $\pm$ 0.04 \dgrunits (60\micron) for a subset of 193 clouds. 
This is significantly lower than the results for the cold LVC and warm LVC populations discussed below.
The stacked images and data are displayed in Figure \ref{fig:main}.
With their unique position-velocity properties outlined in \cite{saul12} and a DGR that agrees with zero, we omit these clouds from the analysis of the rest of
the low-velocity clouds. We discuss the implications of this non-detection in Section \ref{sec:dust_in_hvcs}

\subsection{Low Velocity Clouds}
The warm LVCs have a measured DGR of 0.81 $\pm$ 0.20 \dgrunits (100\micron) and 0.24 $\pm$ 0.06 \dgrunits (60\micron) for 302 clouds 
and the cold LVCs a measured DGR of 0.55 $\pm$ 0.13 \dgrunits (100\micron) and 0.10 $\pm$ 0.04 \dgrunits (60\micron) for 750 clouds.
The results of the bootstrap error analysis are similar to a normal distribution, so we can test for the significance of
the difference in the DGRs of the warm and cold LVC populations by calculating the two sample t-statistic:
\begin{center}
\begin{equation}
t = \frac{(\bar{X_1}-\bar{X_2}) - (\mu_1-\mu_2)}{\sqrt{\frac{s_1}{n_1} + \frac{s_2}{n_2}}} 
\label{equ:tvalue}
\end{equation}
\end{center}
\noindent where $\bar{X_1}, \bar{X_2}$ are the measured means, $\mu_1,\mu_2$ are the actual
means, $s_1,s_2$ are the measured standard deviations, and $n_1,n_2$ are the number of samples (n=1000)
for our bootstrap analysis. We set $(\mu_1-\mu_2)$ to zero to test if the two measured values have a statistically 
non-zero difference. At 100\micron, $t = 34$, and at 60\micron, $t = 19$, both of which have p-values of $<0.0001$,
indicating that the two distributions are distinct. 

\begin{figure}
\includegraphics[scale=0.785,angle=0]{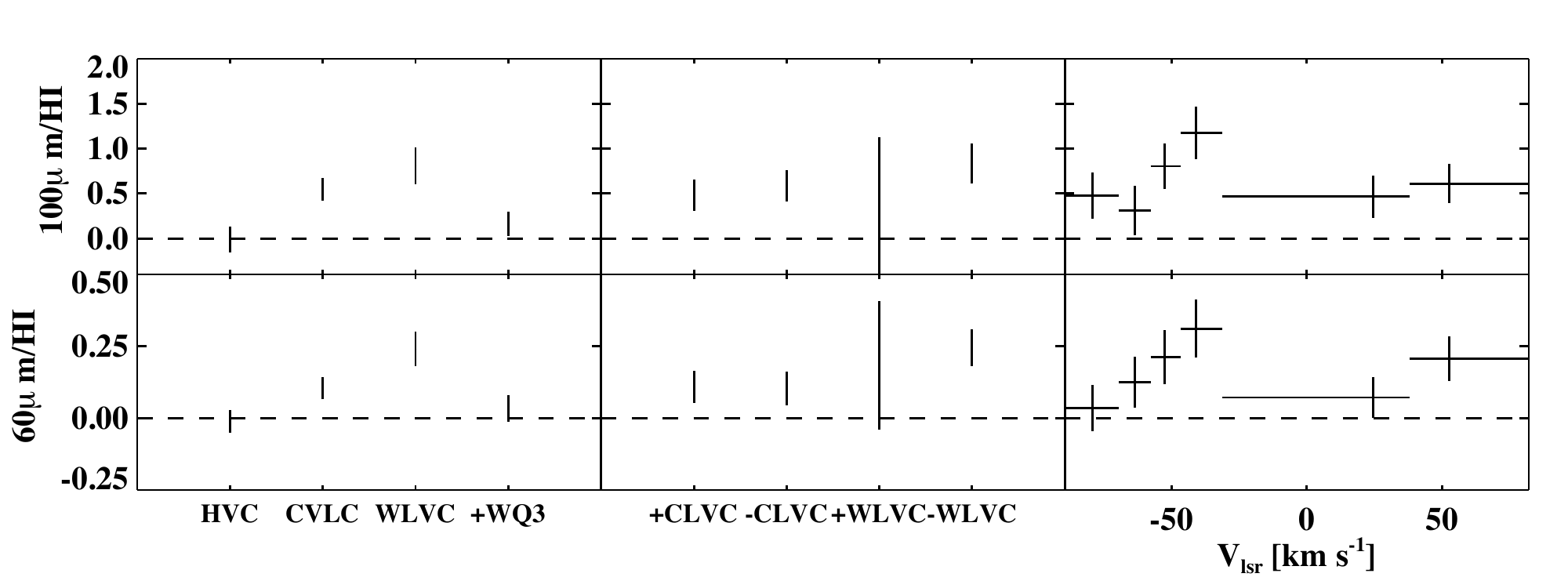}
\includegraphics[scale=0.785,angle=0]{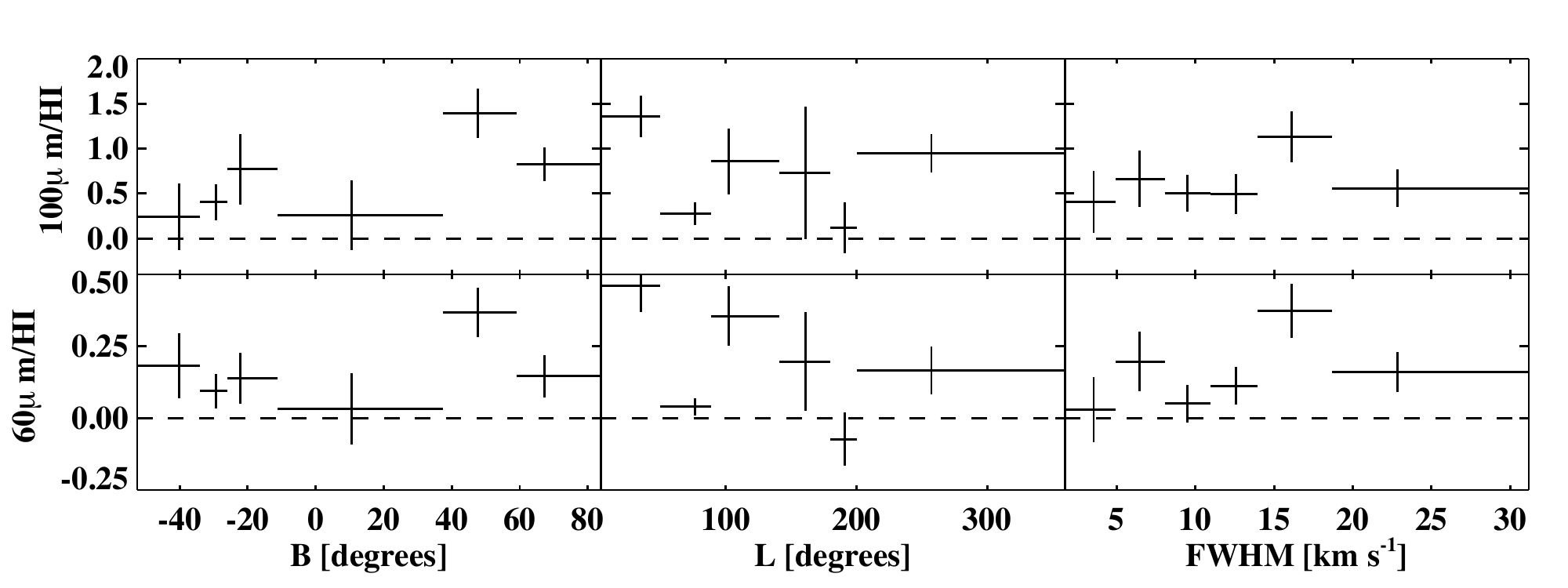}
\includegraphics[scale=0.785,angle=0]{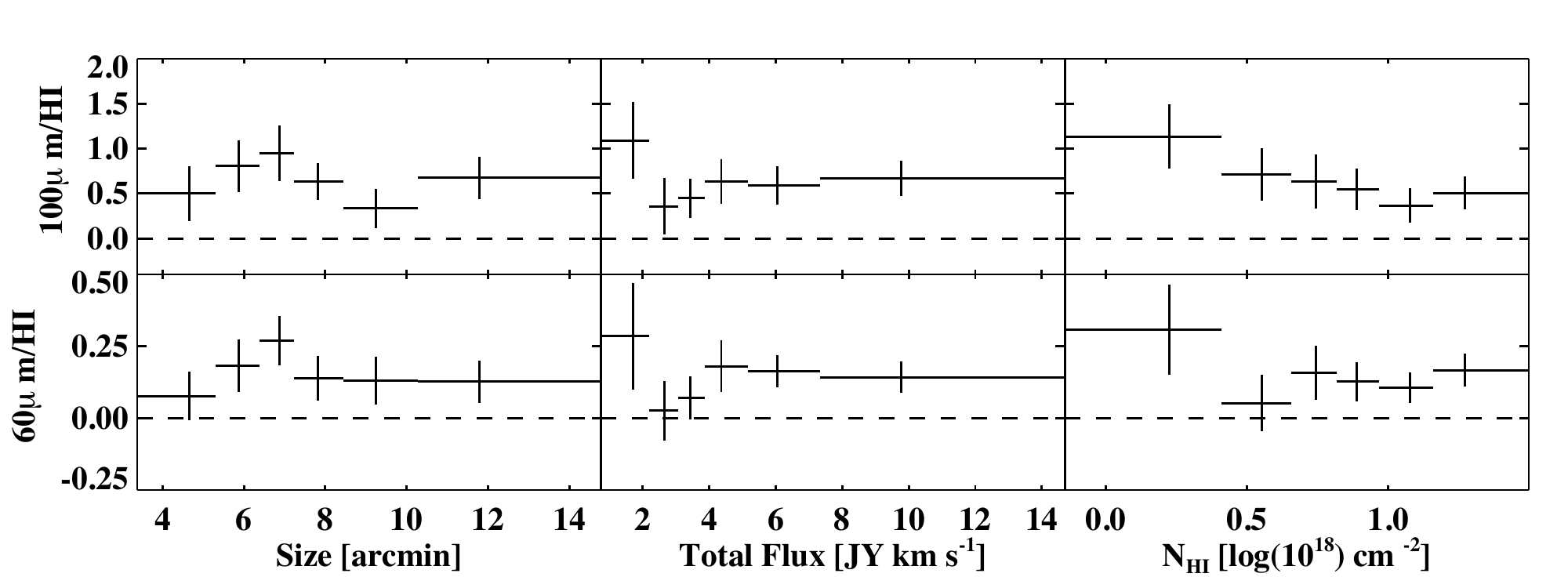}
\caption[100\micron/\HI and 60\micron/\HI Measurements]{
100\micron/\HI and 60\micron/\HI measurements. The upper left plots
show values for each of the four main populations. The upper center plots show just the cold and warm LVCs split into positive and negative LSR velocity bins. The remaining plots show values all the LVCs (cold and warm combined) split by 
velocity, Galactic latitude, Galactic longitude, linewidth, size, total flux, and column density into bins of 175 clouds.}
\label{fig:nine_panel}
\end{figure}

To investigate relationships between the DGR and other parameters, we separate the LVCs into
bins of velocity, linewidth, size, column density, total flux, Galactic latitude, and Galactic longitude. The results are
visualized in Figure \ref{fig:nine_panel}. Each bin has an equal number of clouds (175), so differences in the uncertainties are caused by the data, not sample size.
We do not observe any trends in angular size or total flux.
When separating by column density, we see a downward trend with increasing column density, but only for the
100\micron~measurements. The lower column density bins have increasing uncertainty due to the decreasing total
signal, similar to the total flux plots.
There is a increase in DGR for clouds with moderate negative velocities between -30~\kms and -60~\kms. That velocity range is where the majority of the WLVCs are located. 
In Galactic latitude, we observe non-detections near the Galactic plane ($|b| < 20$) likely caused by increased confusion
with the non-cloud emission. This is echoed in Galactic longitude where the GALFA-HI survey crosses the Galactic plane at $l \simeq 180$ and $l \simeq 60$.

The most interesting relationship is the increased DGR for the linewidth bin near 15~\kms. This is a restatement
of the difference between the warm LVCs and cold LVCs from the beginning of this section. In the 11-14~\kms
~bin we observe a 100\micron ~DGR of 0.49 $\pm$ 0.22 \dgrunits, while in the 14-18.6~\kms ~bin we measure a 100\micron ~DGR of 1.13 $\pm$ 0.28 \dgrunits. 
The DGR for the bin above 19~\kms is consistent with the smaller linewidth bins.

\section{Discussion}
\label{sec:discussion}
\subsection{Non-detections: HVCs and the +WQ3 clouds}
\label{sec:dust_in_hvcs}
As expected, we did not detect dust emission from the HVCs. Two complexes have marginal detections,
but these detections are not enough to warrant discussion beyond what is presented in \S \ref{sec:results_hvcs}.
Previous studies have set lower limits than we are sensitive to \citep{miville05,peek09}, so our non-detections
simply add compact clouds to the body of HVC DGR knowledge.  
There are two explanations for why HVCs are 
not seen in the IR. Either they do not have significant dust, or the dust that is present is not radiating
enough to be detected. A lack of dust is explained by the HVCs being composed of gas that has not 
mixed with Galactic gas, and therefore lacks the processed elements necessary for dust. 
It is possible that HVCs
do contain dust, but that their distance from the Galaxy reduces the heating of the dust by the interstellar
radiation field (ISRF) and therefore the dust does not radiate enough to be detected. In either scenario,
the HVC gas is detectably different from Galactic gas in IR. Using this fact, \cite{peek09}
identify three low-velocity \HI clouds with little or no dust emission that could be low-velocity halo clouds.
Halo clouds at low velocity are too confused with Galactic emission to be easily separated from Galactic disc gas, but a lack
of dust indicates their nature.

We identify the +WQ3 clouds as a population of LVHCs based on their lack of dust emission and the \HI properties
discussed in \cite{saul12}.
The +WQ3 clouds have linewidths greater than 15\kms, which roughly corresponds to the temperature where clouds transition from a fully 
neutral, cold neutral medium to partially-ionized,
warm neutral medium \citep{wolfire03}. Nearly all the HVCs in the GC3 have linewidths greater than 15\kms.
The position-velocity distribution of the +WQ3 clouds agrees with a complex 
of clouds 7-9 kpc from the disc of the Galaxy, with moderate infall velocities (10 - 70\kms). Depending
on the distance, infall velocity, and whether a lagging halo is included, the clouds may or may not be
corotating with the disc. We leave more sophisticated modeling for future work, and submit these clouds as
candidates for further observations to determine their distances and composition. Without more constraints
we cannot say whether these clouds are HVCs with low radial velocities, buoyant recooled clouds as
described by \cite{heitsch09}, or halo clouds forming {\it in situ}, but we can say they are significantly different from the 
rest of the LVCs. The +WQ3 clouds overlap with two of the LVHCs identified by \cite{peek09} (L2 and L5), and we propose that
together they are a large LVHC complex.

\subsection{Detections: Dust in LVCs}
\label{sec:dust_in_lvcs}
Five factors affect the DGR of Galactic gas: metallicity, depletion (fraction of metals in grains), 
ISRF, grain size distribution, and amount of hydrogen in \HI. If we assume that the cold LVCs and warm LVCs are 
subsets of the same population, we do not expect differences in the metallicity, depletion, or ISRF.
It could be argued that the neutral gas column densities of these clouds are somewhat lower than typically 
observed in larger Galactic clouds, and thus the fraction of metals in grains would be reduced, but since the vast majority of refractory elements 
are expected to remain in grains at these column densities, the effect on the grain population is negligible \citep{wakker00}. 
Since we have no indication that the LVCs are far from the Galactic disc, or are composed of material different from the rest of the Galaxy's ISM, 
the ISRF and metallicity are expected to remain constant as well. As clouds become denser, grains can grow through agglomeration, 
which has the effect of increasing the emissivity in the far IR \citep{bernard99,stepnik03}. This is especially true when molecules form at 
low levels in atomic gas \citep[{\eg}][]{planck11_21}. Thus, we might expect the DGR to increase in colder LVCs, which are more likely to harbor denser gas, 
and perhaps even trace molecules. The possibility of a partial transition to molecular hydrogen would also tend to increase our measured 
DGR in colder clouds, as there would be less emission in \HI for a given hydrogen column density. 
Further exacerbating this effect, cold clouds are more likely to be composed of sub-structures that are optically thick to \HI,
weakening the \HI emission and increasing the measured DGR. All these effects have the same sign; 
they increase the DGR in cold LVCs with respect to warm LVCs, 
\emph{contrary to our observations}. 

If instead of assuming the cold clouds are harboring unobserved gas, we assume it is the warm clouds that have an underestimated gas
content, we can account for the increased DGR of some of the warm LVCs. While we don't expect the warm clouds to have molecular gas or self-absorbing
cores, we do know that the warm LVCs have a non-zero ionized fraction \citep{wolfire03}. If we
include ionized gas in the DGR calculation, the DGR of the warm clouds could be lowered to agree with the cold clouds. The
question then becomes: is the dust-to-ionized hydrogen ratio similar to the DGR for neutral gas? If so, the DGR could be used to calculate 
the ionization fraction. The DGR for ionized gas is difficult to measure because the ionized gas is not easily observable \citep{arendt98, schlegel98, lagache99}.

Observations by \cite{peekLeo} of the Local Leo Cold Cloud (LLCC), an extremely nearby (D~$ = 11.3 - 24.3$~pc) \HI cloud, find a
DGR of $0.48$ \dgrunits~after correcting for self-absorption. The LLCC is so cold ($\sim20$K) that it is unlikely that there is any associated ionized gas.
The DGR of the LLCC agrees well with our measurement for cold LVCs. The LLCC measurement is slightly lower than the DGR measured by \cite{schlegel98}
for low density regions, where it is expected that most of the gas would be neutral, but there is significant scatter even for those regions.

Taking the neutral DGR to be $\sim0.5$\dgrunits instead of the canonical $\sim1$\dgrunits of \cite{boulanger88} changes the question
from `why do the cold clouds have a decreased DGR?', to `why are the warm clouds partially ionized?' 
The warm LVCs are predominantly at negative velocities, unlike the cold LVCs which are distributed evenly at positive
and negative velocities. The IVC sky ($\sim 30<|$V$_{lsr}|<90$ \kms) is similarly weighted toward negative velocities \citep{albert04}. Thus, 
we propose that the warm LVCs are associated with the IVCs. This agrees with the results of \cite{planck11} 
who observed higher DGRs and higher dust temperatures in the IVC band than the 
LVC band at both 100\micron
~and 60\micron. 
That we observe higher DGR clouds at negative, but not positive, velocities supports
the model of a supernova-driven galactic fountain where ionized gas is expelled which cools and falls back to the disc \citep{shapiro76, fraternali06, marasco12}.

The observed DGR for clouds with linewidths greater than 19~\kms does not show the same elevated values as the 14-19~\kms bin. This linewith range is the transition between LVCs and HVCs. In GC3, the median HVCs linewidth is 22.0~\kms while the linewidths of the cold and warm LVCs are 8.4 and 20.0~\kms, respectively. The lowered DGR of the $>19$\kms bin relative to the 14-19~\kms bin could be caused by a mixed sample of low DGR, low velocity halo clouds, and high DGR, warm LVCs. 

The cold LVCs are distributed relatively evenly in position at both positive and negative velocities, which indicates
that they are rotating with the disc of the Galaxy, and not too distant as we don't observe the influence of differential Galactic rotation. 
It is possible that the cold LVCs are part of the extremely local ISM like the LLCC. Without distance constraints, it is also possible that
the cold LVCs are at the disc halo interface. The positive velocities do not exclude this possibility as shown by the 
simulations of \cite{heitsch09}. Additional searches for compact cold clouds to build a more complete distribution model are needed as
well as absorption measurements to determine distance constraints.

\section{Conclusion}
\label{sec:conclusion}
We have searched for IR dust emission from the compact neutral hydrogen clouds of the 
GALFA-HI Compact Cloud Catalog using a median image stacking technique and bootstrap error statistics.
We use the IRIS reprocessing of the IRAS 100\micron ~and 60\micron ~bands.
We do not detect dust in the HVC population and while the uncertainty in our measurements are larger than
the previous detections and limits, this is the first constraint for compact clouds as a population.  
We also do not detect dust in the warm, low-positive velocity clouds in the third Galactic quadrant (+WQ3 clouds),
which is additional evidence that these clouds are different from the other LVCs and may be a complex of low-velocity
halo clouds.
We detect dust in the remaining LVCs, with an elevated DGR for clouds with linewidths of 14-19~\kms.  These warm LVCs have a significantly
higher DGR than the cold LVCs that may be due to dust associated with the 
ionized gas component that the cold LVCs may lack.
This result is especially striking because we expected to observe elevated 
dust emission from cold cores in the cold LVCs. The different DGRs of the cold and warm LVCs, along with
the difference in the velocity distributions, suggest that these two populations are distinct.
We propose that the warm LVCs are associated with the warm, infalling IVC populations, while the cold LVCs
are associated with the disc of the Galaxy. 
Finally, we note that the results of this analysis strongly support the population definitions from \cite{saul12}.

\acknowledgements{MEP and DRS acknowledges support from NSF grant AST-0904059, the Luce Foundation and NASA JPL grant 1428009.
Support for this work was provided by NASA through Hubble Fellowship grant HST-HF-51295.01A awarded by the 
Space Telescope Science Institute, which is operated by the Association of Universities for Research in Astronomy, Inc., for NASA, 
under contract NAS 5-26555.}

\clearpage

\clearpage

\bibliographystyle{astron}

\begin{thebibliography}{}

\bibitem[\protect\astroncite{Albert and Danly}{2004}]{albert04}
Albert, C.~E. and Danly, L.: 2004,
\newblock {\em High Velocity Clouds. Edited by Hugo van Woerden} {\bf 312}, 73

\bibitem[\protect\astroncite{Arendt et~al.}{1998}]{arendt98}
Arendt, R.~G., Odegard, N., Weiland, J.~L., Sodroski, T.~J., Hauser, M.~G.,
  Dwek, E., Kelsall, T., Moseley, S.~H., Silverberg, R.~F., Leisawitz, D.,
  Mitchell, K., Reach, W.~T., and Wright, E.~L.: 1998,
\newblock {\em The Astrophysical Journal} {\bf 508}, 74

\bibitem[\protect\astroncite{Beichman}{1987}]{beichman87}
Beichman, C.~A.: 1987,
\newblock {\em IN: Annual review of astronomy and astrophysics. Volume 25
  (A88-13240 03-90). Palo Alto} {\bf 25}, 521

\bibitem[\protect\astroncite{Bernard et~al.}{1999}]{bernard99}
Bernard, J.~P., Abergel, A., Ristorcelli, I., Pajot, F., Torre, J.~P.,
  Boulanger, F., Giard, M., Lagache, G., Serra, G., Lamarre, J.~M., Puget,
  J.~L., Lepeintre, F., and Cambr{\'e}sy, L.: 1999,
\newblock {\em Astronomy and Astrophysics} {\bf 347}, 640

\bibitem[\protect\astroncite{Blitz et~al.}{1990}]{blitz90}
Blitz, L., Bazell, D., and Desert, F.~X.: 1990,
\newblock {\em Astrophysical Journal} {\bf 352}, L13

\bibitem[\protect\astroncite{Boulanger et~al.}{1985}]{boulanger85}
Boulanger, F., Baud, B., and van Albada, G.~D.: 1985,
\newblock {\em Astronomy and Astrophysics (ISSN 0004-6361)} {\bf 144}, L9

\bibitem[\protect\astroncite{Boulanger et~al.}{1998}]{boulanger98}
Boulanger, F., Bronfman, L., Dame, T.~M., and Thaddeus, P.: 1998,
\newblock {\em Astronomy and Astrophysics} {\bf 332}, 273

\bibitem[\protect\astroncite{Boulanger and Perault}{1988}]{boulanger88}
Boulanger, F. and Perault, M.: 1988,
\newblock {\em Astrophysical Journal} {\bf 330}, 964

\bibitem[\protect\astroncite{Collins et~al.}{2009}]{collins09}
Collins, J.~A., Shull, J.~M., and Giroux, M.~L.: 2009,
\newblock {\em The Astrophysical Journal} {\bf 705}, 962

\bibitem[\protect\astroncite{Douglas and Taylor}{2007}]{douglas07}
Douglas, K.~A. and Taylor, A.~R.: 2007,
\newblock {\em The Astrophysical Journal} {\bf 659}, 426

\bibitem[\protect\astroncite{Efron}{1982}]{efron82}
Efron, B.: 1982,
\newblock {\em CBMS-NSF Regional Conference Series in Applied Mathematics}

\bibitem[\protect\astroncite{Fraternali and Binney}{2006}]{fraternali06}
Fraternali, F. and Binney, J.~J.: 2006,
\newblock {\em Monthly Notices of the Royal Astronomical Society} {\bf 366},
  449

\bibitem[\protect\astroncite{Fraternali et~al.}{2013}]{fraternali13}
Fraternali, F., Marasco, A., Marinacci, F., and Binney, J.: 2013,
\newblock {\em The Astrophysical Journal Letters} {\bf 764}, L21

\bibitem[\protect\astroncite{Gott et~al.}{2001}]{gott01}
Gott, J.~R., Vogeley, M.~S., Podariu, S., and Ratra, B.: 2001,
\newblock {\em The Astrophysical Journal} {\bf 549}, 1

\bibitem[\protect\astroncite{Haffner}{2005}]{haffner05}
Haffner, L.~M.: 2005,
\newblock {\em Extra-Planar Gas} {\bf 331}, 25

\bibitem[\protect\astroncite{Heitsch and Putman}{2009}]{heitsch09}
Heitsch, F. and Putman, M.~E.: 2009,
\newblock {\em The Astrophysical Journal} {\bf 698}, 1485

\bibitem[\protect\astroncite{Jones et~al.}{1995}]{jones95}
Jones, M.~H., Rowan-Robinson, M., Branduardi-Raymont, G., Smith, P., Pedlar,
  A., and Willacy, K.: 1995,
\newblock {\em Monthly Notices of the Royal Astronomical Society} {\bf 277},
  1587

\bibitem[\protect\astroncite{Joung et~al.}{2012}]{joung12}
Joung, M.~R., Bryan, G.~L., and Putman, M.~E.: 2012,
\newblock {\em The Astrophysical Journal} {\bf 745}, 148

\bibitem[\protect\astroncite{Lagache et~al.}{1999}]{lagache99}
Lagache, G., Abergel, A., Boulanger, F., D{\'e}sert, F.~X., and Puget, J.-L.:
  1999,
\newblock {\em Astronomy and Astrophysics} {\bf 344}, 322

\bibitem[\protect\astroncite{Lehner et~al.}{2012}]{lehner12}
Lehner, N., Howk, J.~C., Thom, C., Fox, A.~J., Tumlinson, J., Tripp, T.~M., and
  Meiring, J.~D.: 2012,
\newblock {\em Monthly Notices of the Royal Astronomical Society} {\bf 424},
  2896

\bibitem[\protect\astroncite{Maller and Bullock}{2004}]{maller04}
Maller, A.~H. and Bullock, J.~S.: 2004,
\newblock {\em Monthly Notices of the Royal Astronomical Society} {\bf 355},
  694

\bibitem[\protect\astroncite{Marasco et~al.}{2012}]{marasco12}
Marasco, A., Fraternali, F., and Binney, J.~J.: 2012,
\newblock {\em Monthly Notices of the Royal Astronomical Society} {\bf 419},
  1107

\bibitem[\protect\astroncite{Markwardt}{2009}]{mpfit}
Markwardt, C.~B.: 2009,
\newblock {\em Astronomical Data Analysis Software and Systems XVIII ASP
  Conference Series} {\bf 411}, 251

\bibitem[\protect\astroncite{Meyerdierks and Heithausen}{1996}]{meyerdierks96}
Meyerdierks, H. and Heithausen, A.: 1996,
\newblock {\em Astronomy and Astrophysics} {\bf 313}, 929

\bibitem[\protect\astroncite{Miville-Desch{\^e}nes et~al.}{2005}]{miville05}
Miville-Desch{\^e}nes, M.-A., Boulanger, F., Reach, W.~T., and Noriega-Crespo,
  A.: 2005,
\newblock {\em The Astrophysical Journal} {\bf 631}, L57

\bibitem[\protect\astroncite{Miville-Desch{\^e}nes and
  Lagache}{2005}]{miville05iris}
Miville-Desch{\^e}nes, M.-A. and Lagache, G.: 2005,
\newblock {\em The Astrophysical Journal Supplement Series} {\bf 157}, 302

\bibitem[\protect\astroncite{Peek et~al.}{2011a}]{peek11}
Peek, J. E.~G., Heiles, C., Douglas, K.~A., Lee, M.-Y., Grcevich, J.,
  Stanimirovi{\'c}, S., Putman, M.~E., Korpela, E.~J., Gibson, S.~J., Begum,
  A., Saul, D., Robishaw, T., and Kr{\v c}o, M.: 2011a,
\newblock {\em The Astrophysical Journal Supplement} {\bf 194}, 20

\bibitem[\protect\astroncite{Peek et~al.}{2011b}]{peekLeo}
Peek, J. E.~G., Heiles, C., Peek, K. M.~G., Meyer, D.~M., and Lauroesch, J.~T.:
  2011b,
\newblock {\em The Astrophysical Journal} {\bf 735}, 129

\bibitem[\protect\astroncite{Peek et~al.}{2009}]{peek09}
Peek, J. E.~G., Heiles, C., Putman, M.~E., and Douglas, K.: 2009,
\newblock {\em The Astrophysical Journal} {\bf 692}, 827

\bibitem[\protect\astroncite{Peek et~al.}{2007}]{peek07}
Peek, J. E.~G., Putman, M.~E., McKee, C.~F., Heiles, C., and Stanimirovi{\'c},
  S.: 2007,
\newblock {\em The Astrophysical Journal} {\bf 656}, 907

\bibitem[\protect\astroncite{Peek et~al.}{2008}]{peek08}
Peek, J. E.~G., Putman, M.~E., and Sommer-Larsen, J.: 2008,
\newblock {\em The Astrophysical Journal} {\bf 674}, 227

\bibitem[\protect\astroncite{{Planck Collaboration}}{2011a}]{planck11_19}
{Planck Collaboration}: 2011a,
\newblock {\em Astronomy {\&} Astrophysics} {\bf 536}, 19

\bibitem[\protect\astroncite{{Planck Collaboration}}{2011b}]{planck11_21}
{Planck Collaboration}: 2011b,
\newblock {\em Astronomy {\&} Astrophysics} {\bf 536}, 21

\bibitem[\protect\astroncite{{Planck Collaboration}}{2011c}]{planck11}
{Planck Collaboration}: 2011c,
\newblock {\em Astronomy {\&} Astrophysics} {\bf 536}, 24

\bibitem[\protect\astroncite{Putman et~al.}{2012}]{putman12}
Putman, M.~E., Peek, J. E.~G., and Joung, M.~R.: 2012,
\newblock {\em Annual Review of Astronomy and Astrophysics} {\bf 50}, 491

\bibitem[\protect\astroncite{Reach et~al.}{1994}]{reach94}
Reach, W.~T., Koo, B.-C., and Heiles, C.: 1994,
\newblock {\em The Astrophysical Journal} {\bf 429}, 672

\bibitem[\protect\astroncite{Reach et~al.}{1998}]{reach98}
Reach, W.~T., Wall, W.~F., and Odegard, N.: 1998,
\newblock {\em The Astrophysical Journal} {\bf 507}, 507

\bibitem[\protect\astroncite{Saul et~al.}{2012}]{saul12}
Saul, D.~R., Peek, J. E.~G., Grcevich, J., Putman, M.~E., Douglas, K.~A.,
  Korpela, E.~J., Stanimirovi{\'c}, S., Heiles, C., Gibson, S.~J., Lee, M.,
  Begum, A., Brown, A. R.~H., Burkhart, B., Hamden, E.~T., Pingel, N.~M., and
  Tonnesen, S.: 2012,
\newblock {\em The Astrophysical Journal} {\bf 758}, 44

\bibitem[\protect\astroncite{Schlegel et~al.}{1998}]{schlegel98}
Schlegel, D.~J., Finkbeiner, D.~P., and Davis, M.: 1998,
\newblock {\em Astrophysical Journal v.500} {\bf 500}, 525

\bibitem[\protect\astroncite{Shapiro and Field}{1976}]{shapiro76}
Shapiro, P.~R. and Field, G.~B.: 1976,
\newblock {\em Astrophysical Journal} {\bf 205}, 762,
\newblock A{\&}AA ID. AAA017.131.112

\bibitem[\protect\astroncite{Shull et~al.}{2009}]{shull09}
Shull, J.~M., Jones, J.~R., Danforth, C.~W., and Collins, J.~A.: 2009,
\newblock {\em The Astrophysical Journal} {\bf 699}, 754

\bibitem[\protect\astroncite{Sommer-Larsen}{2006}]{sommerLarsen06}
Sommer-Larsen, J.: 2006,
\newblock {\em The Astrophysical Journal} {\bf 644}, L1

\bibitem[\protect\astroncite{Stepnik et~al.}{2003}]{stepnik03}
Stepnik, B., Abergel, A., Bernard, J.-P., Boulanger, F., Cambr{\'e}sy, L.,
  Giard, M., Jones, A.~P., Lagache, G., Lamarre, J.-M., Meny, C., Pajot, F.,
  Peintre, F.~L., Ristorcelli, I., Serra, G., and Torre, J.-P.: 2003,
\newblock {\em Astronomy and Astrophysics} {\bf 398}, 551

\bibitem[\protect\astroncite{Stetson}{1987}]{stetson87}
Stetson, P.~B.: 1987,
\newblock {\em Astronomical Society of the Pacific} {\bf 99}, 191

\bibitem[\protect\astroncite{Thom et~al.}{2006}]{thom06}
Thom, C., Putman, M.~E., Gibson, B.~K., Christlieb, N., Flynn, C., Beers,
  T.~C., Wilhelm, R., and Lee, Y.~S.: 2006,
\newblock {\em The Astrophysical Journal} {\bf 638}, L97

\bibitem[\protect\astroncite{Wakker and Boulanger}{1986}]{wakker86}
Wakker, B.~P. and Boulanger, F.: 1986,
\newblock {\em Astronomy and Astrophysics (ISSN 0004-6361)} {\bf 170}, 84

\bibitem[\protect\astroncite{Wakker and Mathis}{2000}]{wakker00}
Wakker, B.~P. and Mathis, J.~S.: 2000,
\newblock {\em The Astrophysical Journal} {\bf 544}, L107

\bibitem[\protect\astroncite{Wakker and van Woerden}{1997}]{wakker97}
Wakker, B.~P. and van Woerden, H.: 1997,
\newblock {\em Annual Review of Astronomy and Astrophysics} {\bf 35}, 217

\bibitem[\protect\astroncite{White et~al.}{2007}]{white07}
White, R.~L., Helfand, D.~J., Becker, R.~H., Glikman, E., and de~Vries, W.:
  2007,
\newblock {\em The Astrophysical Journal} {\bf 654}, 99

\bibitem[\protect\astroncite{Wolfire et~al.}{2003}]{wolfire03}
Wolfire, M.~G., McKee, C.~F., Hollenbach, D., and Tielens, A. G. G.~M.: 2003,
\newblock {\em The Astrophysical Journal} {\bf 587}, 278

\bibitem[\protect\astroncite{Zech et~al.}{2008}]{zech08}
Zech, W.~F., Lehner, N., Howk, J.~C., Dixon, W. V.~D., and Brown, T.~M.: 2008,
\newblock {\em The Astrophysical Journal} {\bf 679}, 460

\end{thebibliography}

\end{document}